\documentclass[12pt]{iopart}

\usepackage{iopams} 
\usepackage{graphicx}
\usepackage{dcolumn}
\usepackage{bm}
\usepackage{xcolor}
\PassOptionsToPackage{normalem}{ulem}
\usepackage{ulem}
\newcommand{\red}[1]{{\textcolor{black}{#1}}}

\begin{document}

\title{Predicted strong  coupling of solid-state spins via a single  magnon mode}

\author{Denis R. Candido}
 \ead{denisricardocandido@gmail.com}
\address{Department of Physics and Astronomy, University of Iowa, Iowa City, Iowa 52242,
USA}
\address{Pritzker School of Molecular Engineering, University of Chicago, Chicago, Illinois 60637,
USA}
\address{School of Applied and Engineering Physics, Cornell University, Ithaca, New York 14850, USA} 

\author{Gregory D. Fuchs}%
 \ead{gdf9@cornell.edu}
\address{School of Applied and Engineering Physics, Cornell University, Ithaca, New York 14850, USA} 
\address{\red{Kavli Institute at Cornell for Nanoscale Science, Ithaca, New York 14853, USA}}

\author{Ezekiel Johnston-Halperin}
\ead{johnston-halperin.1@osu.edu}
\address{Department of Physics, The Ohio State University, Columbus, Ohio 43210, USA}

\author{Michael E. Flatt\'e}%
\ead{michael\hbox{\_}flatte@mailaps.org}
\address{Department of Physics and Astronomy, University of Iowa, Iowa City, Iowa 52242,
USA}
\address{Pritzker School of Molecular Engineering, University of Chicago, Chicago, Illinois 60637,
USA}
\address{Department of Applied Physics, Eindhoven University of Technology, Eindhoven 5612 AZ, The Netherlands }

\date{\today}


\begin{abstract} 
We propose an approach to realize a hybrid quantum system composed of a diamond nitrogen-vacancy (NV) center spin coupled to a magnon mode of the low-damping, low-moment organic ferrimagnet vanadium tetracyanoethylene. We derive an analytical expression for the spin-magnon cooperativity as a function of  NV  position  under a micron-scale perpendicularly magnetized   disk, and show that,  surprisingly, the cooperativity  will be higher using this magnetic material  than in more conventional materials with larger magnetic moments, due to in part to the reduced demagnetization field.  For reasonable experimental parameters, we predict that the spin-magnon-mode coupling strength is \red{$\textsl{g}\sim 2\pi \times 10$}~kHz.   For isotopically pure $^{12}$C diamond we predict strong coupling of an NV spin to the unoccupied magnon mode, with cooperativity \red{$\mathcal C_\lambda=15$} for a wide range of NV spin locations   within the diamond, well within the spatial precision of  NV center implantation. Thus our proposal describes a practical pathway for single-spin-state-to-single-magnon-occupancy transduction and for entangling NV centers  over micron length scales.
\end{abstract}

\maketitle

Coherent spin centers in solids have emerged as an important solid-state quantum platform due to their applications to quantum sensing~\cite{RevModPhys.89.035002,PhysRevX.10.011003} and  quantum information~\cite{togan2010quantum,de2010universal} as well as their potential for probes of fundamental quantum mechanical properties~\cite{hensen2015loophole}. 
In particular,  negatively-charged nitrogen-vacancy (NV) spin centers in diamond \cite{annurev-conmatphys-030212-184238,doherty2013nitrogen,casola2018probing} are promising candidate{s} for quantum memories and quantum networking~\cite{awschalom2018quantum,Wehnereaam9288} due to their long-lived {spin} coherence combined with  spin-selective optical properties. Creating and scaling entanglement between NV centers poses a key challenge because direct  coupling between them is too weak for separations more than $\sim$20~nm~\cite{neumann2010quantum,dolde2013room,PhysRevB.93.220402}. Independent and simultaneous optical readout of two  NV centers so closely positioned would require heroic efforts due to the spatial and emission-wavelength overlap of their fluorescence.

{Hence alternative schemes to couple two NV centers are of great importance, including heralded entanglement of two NV centers via photons \cite{bernien2013heralded}, as well as the coupling of NV (or other color center) spins to phonons \cite{Habraken2012,ovartchaiyapong2014dynamic,Lemonde2018}.}  {In addition to those, a new} scheme to couple two NV spins proposed the virtual exchange of magnons in ferromagnetic regions separated by a one-dimensional spin chain\cite{lukaprx}. Related work includes  {experimental and theoretical} demonstrations of room-temperature coupling between  NV centers and yttrium iron garnet (YIG) magnon modes in the classical regime~ {\cite{van2015nanometre,page2016optically,du2017control,andrich2017long,kikuchi2017,eichler2017electron,PhysRevB.99.195413}}. 
\begin{figure}[t!]
    \centering
    \includegraphics[width=1\linewidth]{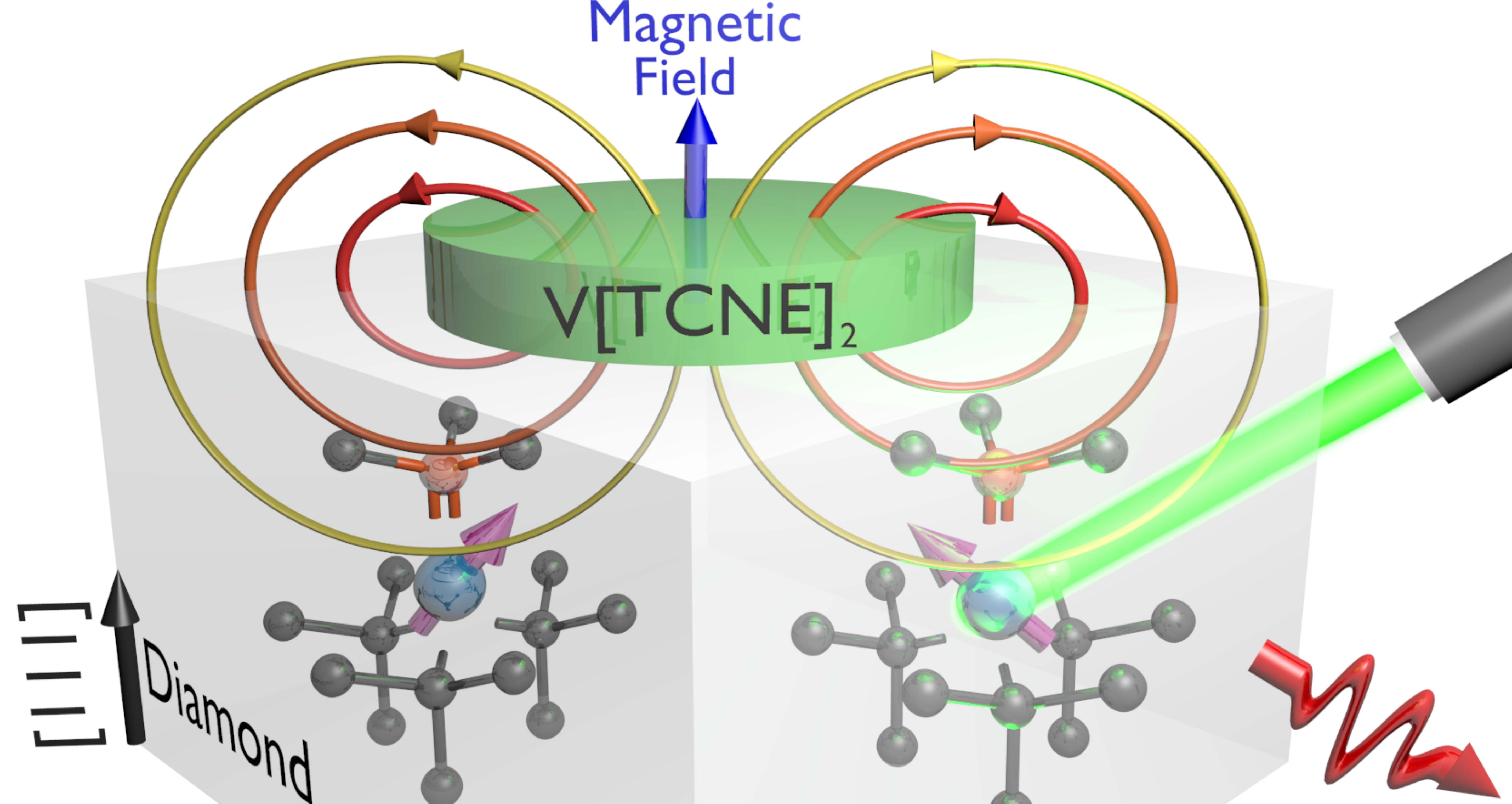}
    \caption{Schematic view of the strong quantum-coherent coupling between  NV-center spin and magnon mode. The green disk represents the normally magnetized {V[TCNE]$_x$} ferrimagnetic material placed on top of a diamond [111] substrate possessing  NV centers. }
    \label{fig1}
\end{figure}
 However, while YIG  exhibits ultra-low magnetic dissipation, integrating {this material} with diamond poses significant hurdles. First, direct epitaxy on diamond has not been demonstrated, likely because of the substantial lattice mismatch. Second, YIG films exhibit a sizable increase in magnon damping both when they are patterned~\cite{hahn2014measurement,jungfleisch2015spin} and at  low temperature~\cite{haidar2015thickness,jermain2017increased} --- both of which will be necessary to obtain strong interactions and low thermal magnon occupation. Recent efforts to address the first two points with a high temperature (800$^{\circ}$C) post-growth  annealing~\cite{Heyroth2019} show promise in restoring intrinsic material properties, but present their own challenges for device integration.

Here we {propose} 
a practical approach to engineer strong, coherent single-spin-single-magnon{-mode} (hereafter spin-magnon) coupling, which would permit transduction between the spin state of a single NV \red{and} the magnon occupancy of an otherwise unoccupied magnon mode.  This hybrid quantum system can enable entanglement between NV centers separated by $\approx 1$~$\mu$m. Our proposal relies on the magnetic properties of the low loss organic ferrimagnetic material  vanadium tetracyanoethylene  \cite{pokhodnya2013carrier,yu2014ultra,cimpoesu2014disorder,harberts2015chemical,zhu2016low,liu2018organic,chilcote2019spin,franson2019low} (V[TCNE]$_x$) to overcome the {challenges} of building and integrating low loss magnetic materials on diamond. V[TCNE]$_x$ can be grown without lattice-matched substrates and  patterned at the micron scale into arbitrary shapes while maintaining its low magnetic damping\cite{franson2019low}  {(Gilbert parameter $\alpha = 4\times 10^{-5}$, comparable
to very high-quality  YIG, and magnon damping rate} \red{$\mathsf{\varkappa}=2\pi \times 100$~kHz)}. Our proposal considers the resonant spin-magnon 
coupling \red{($\textsl{g}$)} achievable at low temperature ($T\lesssim 100$~mK) with a micron-scale disk of perpendicularly-magnetized {V[TCNE]$_x$} positioned on top of a single-crystal  (111) diamond substrate with  {previously-}embedded NV centers  (Fig.~\ref{fig1}). 
Our analytical calculations predict  \red{$ \textsl{g} \sim 2\pi \times 10$~kHz} between an NV spin 30~nm below the diamond surface and an unoccupied magnon mode of a micron-diameter, 100~nm thick disk. The strong in-plane fringe fields  of the V[TCNE]$_x$ microstructure efficiently couple to an NV spin with axis perpendicular to the surface in a (111) diamond substrate\cite{Fuchs2011}.  {The cooperativity {~\cite{RevModPhys.86.1391}},
\begin{equation}
\red{{\mathcal C}_\lambda=\frac{4{\textsl{g}}_{\lambda}^2}{n{\mathsf{\varkappa}}/{T_{2}^{*}}},}
\label{coop}
\end{equation}
characterizes how often the magnon-spin system can coherently swap quantum information between  magnon occupancy and the spin state. 
Here $T_{2}^*$ is the free precession coherence time of  NV center, and $n=1/\left(e^{{\hbar \omega_\lambda}/{k_b T}}-1\right)$  the average thermal number of magnons, where $k_B$ is the Boltzmann constant and $\hbar \omega_\lambda$ is the magnon energy. We assume\cite{t2star} NV coherence time $T_{2}^{*}\approx 1.5\times 10^{-3}$~s that is obtainable\cite{t2star} in isotopically-pure $^{12}$C  diamond and note that for $T\lesssim 100$~mK,  $n\approx 1$, yielding cooperativity } \red{$\mathcal{C}\approx 15$}  {for a range of  NV center positions under the disk.}
This strong spin-magnon coupling, when combined with suitable  gates to tune the system into and out of resonance, can be used to mediate controllable coupling between multiple NV centers separated from each other by microns but each coupled to the magnon mode. We further obtain an analytical formula for the spin-magnon cooperativity expressed as a product of the magnon frequency, a geometric \red{factor}, and a material factor. This allows us to compare the cooperativity for systems  with different disk dimensions, as well as for different materials realized with the same geometry. Using this formula we predict a lower cooperativity for thin-film YIG compared to {V[TCNE]$_x$}, due to  (1) the higher magnetic damping, and (2) the much higher resonance energy for a YIG magnon due to its much larger demagnetization field.  {We conclude with a discussion of the (substantial) experimental challenges that must be overcome to implement this proposal.}

{\it Magnons ---} The spin-magnon coupling $\red{\textsl{g}}$ will be enhanced for high angular momentum magnon modes of a perpendicularly-magnetized disk ($\textbf{B}_{dc}=\mu_0 \textbf{H}_{dc}=B_{dc} \hat{z}$), which will localize the magnon mode near the edge of the disk.
Analytic expressions for the magnon mode frequencies ($\omega$), rf magnetization ($\textbf{m}$) and fringe fields ($\red{{\textbf{h}}}$) for a disk of {V[TCNE]$_x$} emerge from  Maxwell's equations in the magnetostatic limit\cite{magnetostatic} together with the Landau-Lifshitz-Gilbert equation {\cite{landau1992theory,gilbert1955equation}}, 
\begin{eqnarray}
&&\boldsymbol{\nabla}\times\textbf{H}  =0,\label{rotH}\\
&&\boldsymbol{\nabla}\cdot\textbf{B}  =0,\label{divB}\\
\frac{d\textbf{M}}{dt}+\gamma\textbf{M}\times\textbf{H}&=& - \alpha(\gamma/M_S) \textbf{M}\times\left(\textbf{M}\times \textbf{H} \right),\label{LL}
\end{eqnarray}
with the magnetic field $\textbf{H}=({\textbf{B}}/{\mu_0})-\textbf{M}$. Here  $\textbf{M}$ is the magnetization, $\mu_0$ is the vacuum permeability,  $\gamma$ is the \red{magnet's} gyromagnetic ratio, $\alpha$ is the  damping factor and $M_S$ is the saturation magnetization (see Supplemental Material).
 The disk occupies the region $\mathcal{D}=\left\{\left(r,\theta,z\right),\thinspace r<R\thinspace\thinspace\&\thinspace\thinspace\left|z\right|<d/2\right\}$ and the substrate $\mathcal{S}=\left\{\left(r,\theta,z\right),\thinspace\thinspace z<-d/2\right\}$ (Fig.~\ref{fig1}). 
\begin{figure}[t!]
    \centering
    \includegraphics[width=.7\linewidth]{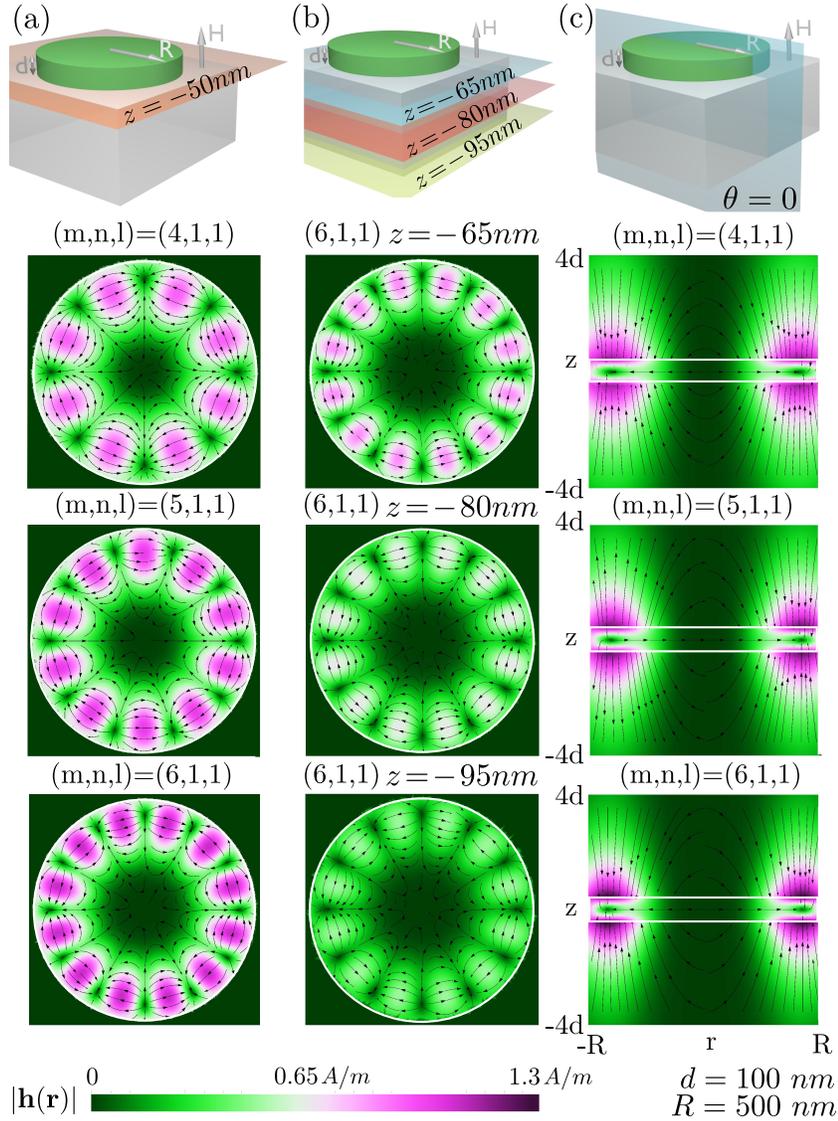}
    \caption{(a) In plane fringe fields $\textbf{h}(\textbf{r})$ at  {$z=-50$}~nm (orange plane) for the modes $(4,1,1)$, $(5,1,1)$, $(6,1,1)$. (b) In plane fringe fields $\textbf{h}(\textbf{r})$ for the mode $(6,1,1)$ at depth  {$z=-65$}~nm,  {$z=-80$}~nm and  {$z=-95$}~nm (blue, red and yellow planes, respectively). (c) Fringe fields $\textbf{h}(\textbf{r})$ for cross section view $\theta=0$ (blue plane) for modes $(4,1,1)$, $(5,1,1)$, $(6,1,1)$. The white lines delimit the disk dimensions $R=500$~nm and $d=100$~nm.}
    \label{fig2}
\end{figure}
 {Assuming pinning of the magnetization at the edge of the disk~\cite{sparks1970,sparks1970-ssc}, we obtain spin waves with frequencies (Supplemental material)}

\begin{equation}
\omega_{\lambda}=\gamma\left\{ \tilde{H}_{0}^{2}+\frac{\tilde{H}_{0}M_{S}}{1+{{\red{k_{i,\lambda}^2}}}/\red{{k_{o,\lambda}^{2}}}}\right\}^{1/2},
\label{magnon-f}
\end{equation}
where the mode index $\lambda=(m,n,l)$ stands for angular ($m)$, radial ($n$) and thickness ($l$) modes\red{, $k_{i,\lambda}$ and $k_{o,\lambda}$ are, respectively, the thickness and radial magnon wave numbers, with $k_{o,\lambda}={\beta_{m-1}^n}/{R}$, where $\beta_{m-1}^{n}$ is the n-th zero of $J_{m-1}(k_{o,\lambda} R)$}. The analytic magnon frequencies obtained here [Eq.~(\ref{magnon-f})] were experimentally and numerically validated for  V[TCNE]$_x$ disks~\cite{franson2019low}.
The fringe field $\textbf{h}_\lambda (\textbf{r},t)=\textrm{Re}\left\{\bar{\textbf{h}}_\lambda(\textbf{r})\red{e^{-i\omega_\lambda t}} \right\}$ \red{(where $\bar{\textbf{h}}_\lambda(\textbf{r})$ is a complex coefficient)}  within the ${\mathcal S}$ region 
\begin{eqnarray}
\frac{\textbf{h}_{\lambda}\left(\textbf{r},t\right)}{\red{m_{0,\lambda}}} && =  {\hat{r}\cos\left(m\theta\red{-\omega_\lambda t}\right)\left[\red{k_{o,\lambda}J_{m-1}\left(k_{o,\lambda}r\right)-\frac{m}{r}J_{m}\left(k_{o,\lambda}r\right)}\right]e^{\red{k_{o,\lambda}}z}}\nonumber\\
&& -\bigg[\frac{m\hat{\theta}}{r}\sin\left(m\theta\red{-\omega_\lambda t}\right)\red{-\hat{z}k_{o,\lambda}}\cos\left(m\theta\red{-\omega_\lambda t}\right)\bigg]J_{m}\left(\red{k_{o,\lambda}}r\right)e^{\red{k_{o,\lambda}}z} \label{fring}
\end{eqnarray}
\red{where} $\red{m_{0,\lambda}}=\left({{2 \gamma \hbar M_{S}}/{\red{2\pi}k_{o,\lambda}^{2}\kappa^{2}I_{\lambda}^{r}\red{I_{\lambda}^{z}}}}\right)^{1/2}$
with $I_{\lambda}^{r}=\int_{0}^{R}rdrJ_{m-1}^{2}\left(k_{o,\lambda}r\right)$ \red{the normalization factor for the radial dependence of the magnon} and $\red{I_{\lambda}^{z}}$ \red{the normalization factor for the height dependence of the magnon (Supplemental material).}

Figure~\ref{fig2} presents the fringe field $\textbf{h}_{\lambda}(\textbf{r},t)$ [Eq.~(\ref{fring})] for a {V[TCNE]$_x$} disk with $d=100$~nm,  {$R=500$~nm}, $M_S=7560$~A/m~\cite{yu2014ultra}, $A_{ex}=2.2\times 10^{-15}$~J/m~\cite{franson2019low}, $ {{\gamma}/{2\pi}}=({4\pi}/{1000})\times 2.73\times 10^6$~Hz m/A,~\cite{franson2019low} and $B=B_c\approx 56$~mT. The first column [Fig.~\ref{fig2}(a)] displays the in plane $\textbf{h}_{\lambda}(\textbf{r},t)$ profile for three different modes $(m,n,l)$ at the interface  {$z=-50$}~nm between the disk $\mathcal{D}$ and the diamond substrate $\mathcal{S}$ [orange plane Fig.~\ref{fig2} (a)]. The largest fringe fields occur for the lowest thickness and radial modes\cite{fringe} $l=1$ and $n=1$. 
The larger the magnon angular index $m$, the more localized the fringe field around the edge of the disk and the larger the amplitude of the fringe fields. Therefore, high angular index magnon modes are preferred for  coupling to  NV center spins.

The fringe field is largest  near the interface between the magnetic disk and the diamond, therefore producing the largest values of \red{$\textsl{g}$}. However,  the coherence time for an NV center spin  decreases sharply as it approaches the surface due to  magnetic noise from uncontrolled surface spins~\cite{nvdepth1,nvdepth2,nvdepth3,nvdepth4}. Recent work has shown that the spin coherence time saturates to the bulk value for   depths of $\approx 30$~nm~\cite{nvdepth3}. For this reason, we focus on NV centers positioned  a minimum of $30$~nm below the surface. Figure~\ref{fig2}(b) depicts the decay of the in-plane fringe field of the  $(6,1,1)$ mode into the diamond substrate, showing the field at the depths of  {$z=-65$}~nm,  {$z=-80$}~nm and  {$z=-95$}~nm represented by the blue, red and yellow planes, respectively. Although the fringe field decays exponentially into the diamond  {substrate}, $|\textbf{h}| \propto e^{\red{k_{o,\lambda}}z }$ [Eq.~(\ref{fring})], the attenuation of the field amplitude from  {$z=-65$}~nm to  {$z=-95$}~nm is only {roughly} a factor of $1/2$. In the thin disk regime, $R\gg d$,  $|\textbf{h}|$  decays with a scale $\approx R$ and not $\approx d$, as $e^{\red{k_{o,\lambda}} z} \approx 1 + z/R$. The enhancement and localization of the fringe field amplitude for larger angular modes $m$ can also be visualized in Fig.~\ref{fig2}(c), where the fringe fields are shown in a cross section plane $\theta=0$ [blue plane in Fig.~\ref{fig2}(c)] for the modes $(4,1,1)$, $(5,1,1)$ and $(6,1,1)$. 
\begin{figure}[t!]
    \centering
    \includegraphics[width=.7\linewidth]{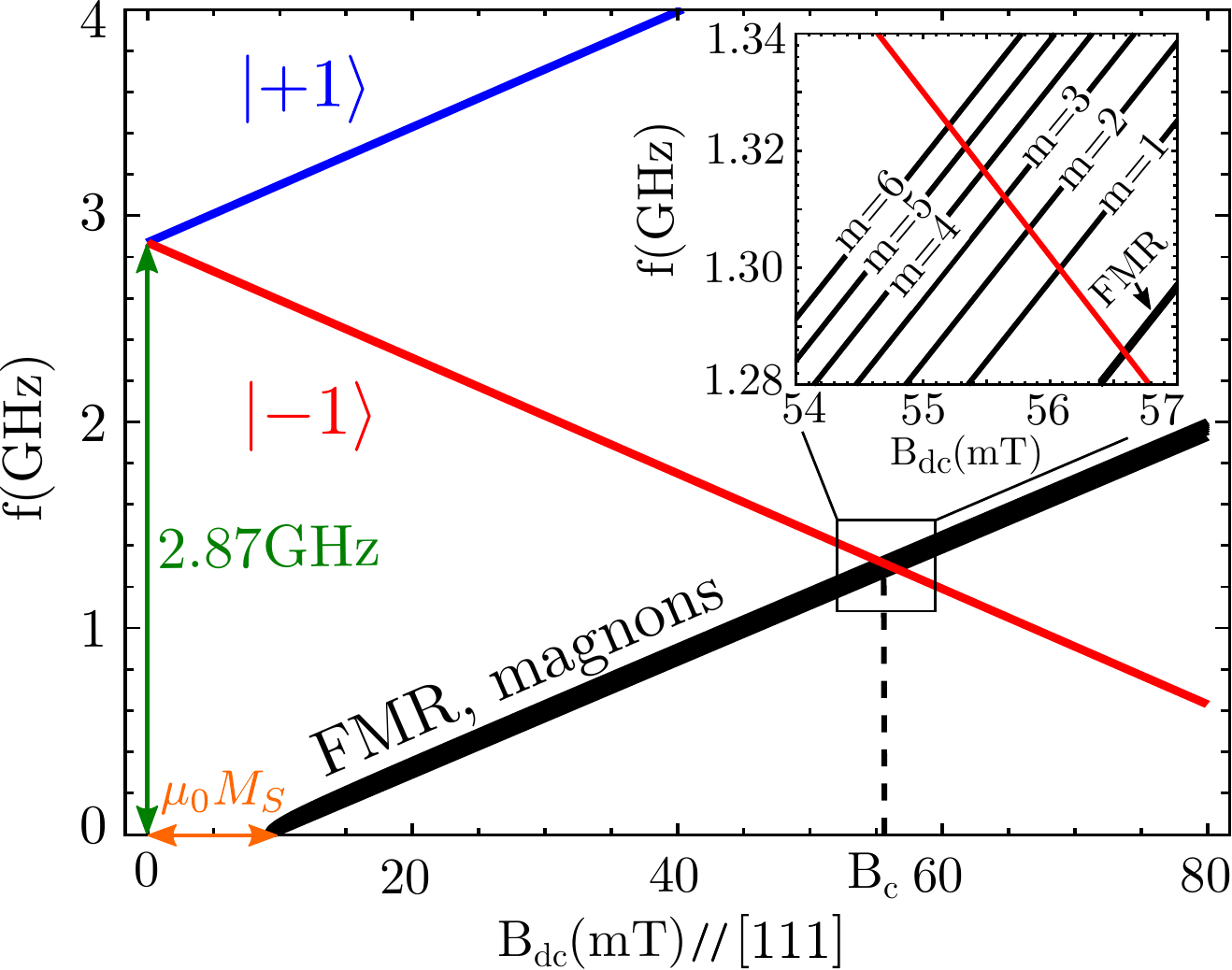}
    \caption{Frequencies of  NV center levels $\left|\pm 1\right\rangle$ (blue and red solid lines), FMR and magnons (black solid lines) as a function of external dc magnetic field {  {$B_{dc}$} parallel to the [111] Diamond crystallographic direction} and NV center axis. Inset shows a zoom-in of the crossing region between $\left|-1\right\rangle$ level with both magnonic $m=1,2,3,4,5,6$ and FMR frequencies.}
    \label{fig3}
\end{figure}

{{\it  Spin-magnon coupling ---}} Here we investigate the coupling between magnon {modes} and NV center spins. In the presence of an external magnetic field $\textbf{B}_{ex}$, the ground state spin 1 Hamiltonian for an NV center, written in the $\left|m_s\right\rangle$ basis along the NV axis, $\left|-1\right\rangle ,\left|0\right\rangle ,\left|+1\right\rangle $ is~\cite{loubser1978electron,van1990electric,tamarat2006stark,hossain2008ab,de2010universal,togan2010quantum,bassett2011electrical,maze2011,dolde2011,doherty2011negatively,acosta2012dynamic,doherty2012,doherty2013nitrogen,dolde2014,schirhagl2014nitrogen,dolde2014nanoscale,rogers2015singlet,ivady2015theoretical,christle2017,anderson2019electrical}
\begin{equation}
\red{\frac{{\cal H}_{\red{\rm NV}}}{2\pi\hbar}  =  D\left(S_{z}^{2}-\frac{2}{3}\right)+\frac{\gamma_{\red{\rm NV}}}{ 2\pi}\textbf{B}_{ex}\cdot\textbf{S}}
\label{nvcenter-f}
\end{equation}
where $\red{\gamma_{\red{\rm NV}}}$ is the \red{NV-center gyromagnetic ratio}, $\textbf{S}$ is represented by the spin-1 matrices, $\hbar$ is the reduced Planck constant and $D=2.87$~GHz is the zero-field splitting between $\left|0\right\rangle $ and the degenerate states at zero field $\left|\pm 1\right\rangle $. 

In Fig.~\ref{fig3} we plot the resonant frequencies of the uniform  {ferromagnetic main resonance (FMR)} mode, magnon modes, [Eq.~(\ref{magnon-f})] and the NV center spin [Eq.~(\ref{nvcenter-f})] as a function of
$B_{dc}$
oriented along the {[}111{]} symmetry axis of the NV center (Fig.~\ref{fig1}). The frequency corresponding to the \red{$\left|0\right\rangle \longleftrightarrow \left|-1\right\rangle$} transition crosses both the FMR and magnon frequencies at $B_c \approx 56$~mT and $f\approx 1.30$~GHz. Due to the low saturation magnetization of {V[TCNE]$_x$}, the different angular mode frequencies associated \red{with} the magnon modes $m=1,2,\red{3},4,5,6$ are not resolved within the scale of Fig.~\ref{fig3}. The inset of Fig.~\ref{fig3} provides enough resolution {to see} that the different angular modes $m=1,2,\red{3},4,5,6$ and the FMR can now be distinguished.

 {Here we show that the spin-magnon coupling Hamiltonian \cite{lukaprx}} can be transformed into a Jaynes-Cumming Hamiltonian~\cite{jaynes1963}, in which the coupling originates from the Zeeman Hamiltonian \red{$\hbar \gamma_{\red{\rm NV}}  \textbf{B}_{ex}\cdot \textbf{S}$}. The total magnetic field is a sum of the dc applied external magnetic field $\textbf{B}_{dc}$ and the magnon rf fringe fields $\textbf{h}_{\lambda}(\textbf{r},t)$ [Eq.~(\ref{fring})].
After applying the rotating wave approximation (RWA), and under resonant conditions, the NV center-magnon coupling Hamiltonian  {(Supplemental Material)}

\begin{equation}
\red{{ {\cal H}_I  =\hbar \textsl{g}_{\lambda}^{\star}(\textbf{r}_{\red{\rm NV}})a_{\lambda}^{\dagger}{\sigma}_{+}^{D}+\hbar\textsl{g}_{\lambda}(\textbf{r}_\red{\rm NV})a_{\lambda}{\sigma}_{-}^{D},}}
\label{Hint}
\end{equation}
{with} \red{ $\sigma_{+}^{D}\left|-1\right\rangle =\left|0\right\rangle $, $\sigma_{-}^{D}\left|0\right\rangle =\left|-1\right\rangle $ and} coupling between the NV center spin and the $\lambda$ magnon mode 
\red{
\begin{eqnarray}
\mathcal{\textsl{g}}_{\lambda}\left(\textbf{r}_{\red{\rm NV}}\right) & = & \frac{\sqrt{2}}{4}\gamma_{{\rm NV}}\mu_{0}\thinspace\bar{h}_{+,\lambda}(\textbf{r}_{\red{\rm NV}})        \label{gcoup} \\
 & = & \frac{\sqrt{2}}{4}\gamma_{\red{\rm NV}}\mu_{0}\thinspace m_{0,\lambda}\left[k_{o,\lambda}J_{m-1}\left(k_{o,\lambda}r_{\red{\rm NV}}\right)-2\frac{m}{r}J_{m}\left(k_{o,\lambda}r_{\red{\rm NV}}\right)\right]
 e^{im\theta_{\rm NV}}\red{e^{{k_{o,\lambda}}z_{\rm NV}}} \nonumber
\end{eqnarray}}
for a NV center located at $\textbf{r}=\textbf{r}_\red{\rm NV}=\red{(r_{\red{\rm NV}},\theta_{\red{\rm NV}},z_{\red{\rm NV}})}$. \red{This shows that the transition $\left|0\right\rangle \longleftrightarrow\left|-1\right\rangle$ is only allowed by the left circularly polarized component of the fringe field, i.e., $\bar{h}_{+,\lambda}(\textbf{r})$.}
Here we write  {Eq.~(\ref{gcoup})} to preserve the position-dependent phase of $ \textsl{g}_\lambda$ to account for the possibility of coupling two NV spins to the same magnon mode.
In Eq.~(\ref{Hint})  the term \red{$a_{\lambda}^{\dagger}\sigma_{+}^D$ ($a_{\lambda}\sigma_{-}^D$)} corresponds to the raising (lowering) of the NV center spin, combined with a creation   (annihilation) of a magnon (since the $\left|-1\right\rangle$ state has higher energy than the $\left|0\right\rangle$ spin state). 
For the magnon mode $\lambda=(6,1,1)$ at $30$~nm below the bottom disk surface  {($z=-80$~nm)}, we find \red{$|\bar{h}_{+,\lambda}| \approx 0.6$~A/m} [See Fig.~\ref{fig2}(b)], which yields \red{$\textsl{g}_{(6,1,1)}=2\pi \times 7.7$}~kHz. 
The magnon damping rate \red{$\mathsf{\varkappa} = 2\pi\times 100$}~kHz is obtained from the homogeneous ferromagnetic resonance linewidth, \red{$\mathsf{\varkappa} = 2\pi f/Q$}, where\cite{rossing1963resonance} $Q=1/(2\alpha)$ with\cite{franson2019low} $\alpha=4\times 10^{-5}$ and $f\approx 1.3$~GHz for {V[TCNE]$_x$}.\cite{temper} 
As an upper bound,  \red{$\mathsf{\varkappa}\sim 2\pi \times 300$}~kHz from the inhomogeneous broadening linewidth, for which\cite{franson2019low} $Q=3700$ {.} 

 {We calculate the cooperativity Eq.~(\ref{coop}), assuming a $T_{2}^{*}\approx 1.5\times 10^{-3}$~s  that is  constant with respect to the defect depth; defects  deeper than $\approx 30$~nm have bulk-like $T_2^*$'s\cite{nvdepth3,nvdepth4}.}
\begin{figure}[t!]
    \centering
    \includegraphics[width=.7\linewidth]{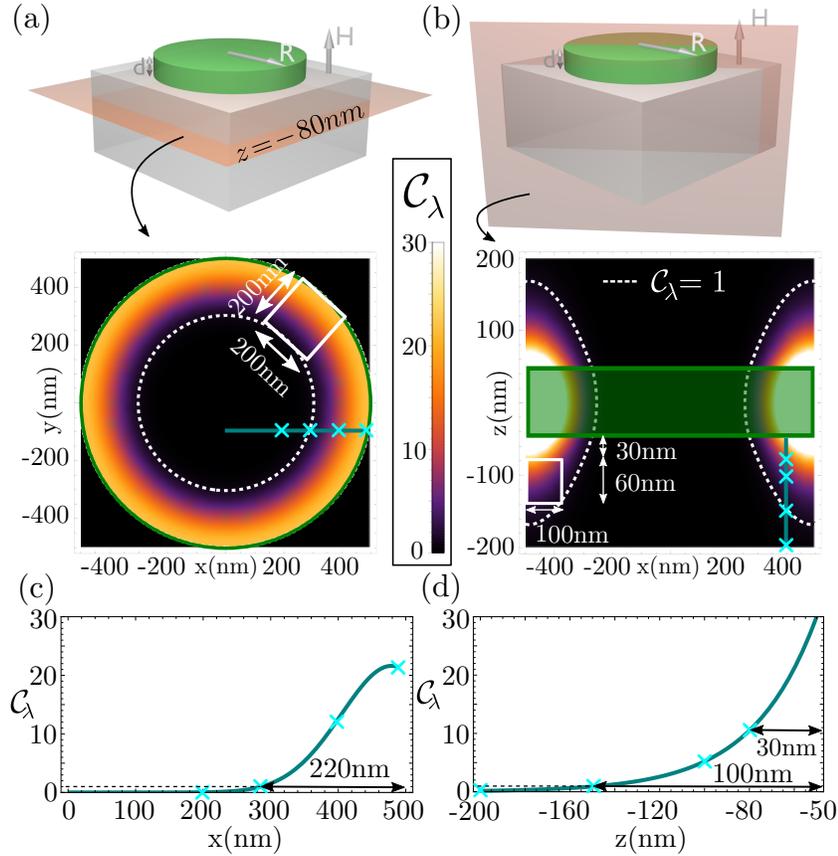}
    \caption{\red{Spatial plot} of the cooperativity of the $\lambda=(6,1,1)$ magnon mode (a) at $30$~nm below the disk  {($z=-80$~nm)}, (b) within cross-section plane, (c) along the teal line within Fig.~\ref{fig4}(a), (d) along the teal line within Fig.~\ref{fig4}(b). The dashed white border shows the strong-\red{coupling} regime stability region where ${\mathcal C}_\lambda \geq 1$, where the white rectangle indicates a tolerance for spatial implantation imprecision for the NV centers while still achieving high cooperativity. The green lines delimit the disk dimension $d=100$~nm and $R=500$~nm.}
    \label{fig4}
\end{figure}
We find that the cooperativity  {${\mathcal C}_\lambda>1$} for a wide range of positions of the NV spin near the disk, as shown in Fig.~\ref{fig4} for the  {$\lambda=({6,1,1})$} magnon mode.  Figure~\ref{fig4}(a) shows  {$\mathcal C_\lambda$} for the plane positioned $30$~nm below the disk  {($z=-80$~nm)}, whereas Fig.~\ref{fig4}(b) shows  {$\mathcal C_\lambda$} as a function of the  NV depth in a cross-sectional plane.   {${\mathcal C}_\lambda>1$} within those regions surrounded by white dashed lines in Fig.~\ref{fig4}(b). We focus on the regions delineated by solid white lines, whose dimensions are $\ge 100$~nm,
 because the spatial precision of implanting  NV spins is $\approx 20$nm, due to straggle and finite aperture size~\cite{LeeNV,McLellan2016}. 
Figs.~\ref{fig4}(c) and (d) show the variation of the cooperativity along the teal lines in Figs.~\ref{fig4}(a) and (b), respectively. Therefore, two  NV centers that in principle could be directly coupled only if separated by $\lesssim 20$~nm, could now be coupled through a magnon mode when separated by $\approx 2R \approx 1$~$\mu$m.

The cooperativity, Eq.~(\ref{coop}), is the product of a geometric factor $\Gamma_{\lambda}\left(R,d,\textbf{r}\right)$ (dimension and position dependent), the reciprocal magnon energy $\hbar \omega_\lambda$ and \red{a material-dependent factor $M_S/\alpha$, } 
\begin{equation}
\red{\mathcal{C}_\lambda =\frac{\gamma}{2\pi}{ \frac{T_{2}^{*}}{n}\left({\hbar \gamma_{\red{\rm NV}}\mu_{0}}{}\right)^{2}} \times \Gamma_{\lambda}\left(R,d,\textbf{r}_{\red{\rm NV}}\right)\times\frac{1}{\hbar\omega_{\lambda}}\times\frac{M_{S}}{\alpha}.\label{coop-factor} }
\end{equation}
The linear dependence on $M_S$ comes from the amplitude of the fringe fields. 
This equation
provides an analytical formula for the cooperativity of any magnetic disk on top of a diamond substrate, and thus it, along with  estimates for the quantities therein, are significant theoretical results of our work.
Using this formula, we can analytically compare  the cooperativity for { the} same material {with} different {disk dimensions}, and  for different materials in the same geometry.  For instance, we evaluate the cooperativity for the same magnon mode $(6,1,1)$ for a disk with double the diameter ($R=1$~$\mu$m). We obtain through Eq.~(\ref{coop-factor}) that \red{$C_{(6,1,1)}^{1\mu m}\approx 4.5$} for a NV center spin located $30$~nm below the V[TCNE]$_x$ bottom surface; { in general,} disks with smaller diameter than \red{$\approx 4~\mu$m} will permit $C_{(6,1,1)}>1$ for this magnon mode and NV depth.

{\it Comparison with YIG.---}  
For out-of-plane magnetization there is {no resonance} between the above magnon mode in a YIG disk  and  NV center levels, considering low-temperature thin film values\cite{d2013inverse,haidar2015thickness,jermain2017increased} of $M_S$ ($\approx 2200$~G)~(see Supplemental Material). The shift by $\mu_0 M_S$ of both magnon and FMR frequencies along the $B$ axis [coming from $f\approx \gamma(B/\mu_0 - M_S)$] moves the magnonic and FMR dispersions away from the  NV level (see orange arrow within Fig.~\ref{fig3}). A resonance condition occurs for $M_S\lesssim 1800$~\red{G} or for much higher index magnon modes. \red{We also note that for YIG, the Hamiltonian [Eq.~(\ref{Hint})], coupling [Eq.~(\ref{gcoup})] and cooperativity [Eq.~(\ref{coop-factor})] expressions assume a different form (see Supplemental Material) due to coupling to the right circular polarization of the fringe field. This occurs because the resonance occurs for magnetic fields which switch the energies of the $\left|0\right\rangle$ and $ \left|-1\right\rangle$ states.} Furthermore, at low temperature the damping  for thin-film YIG increases by a factor of $\approx 6-30$, reaching 
$\approx 1.5\times 10^{-3}$ in Ref.~\cite{haidar2015thickness} and 
$\approx 18\times 10^{-3}$ in Ref.~\cite{jermain2017increased} --- possibly due to the impurity relaxation from rare earth or  Fe$^{2+}$ ions~\cite{PhysRevLett.3.30,PhysRevLett.3.32,doi:10.1063/1.1984756,PhysRev.123.1937,PhysRev.133.A728,doi:10.1063/1.1984589}. Using now Eq.~(\ref{coop-factor}) we compare the cooperativity between YIG and {V[TCNE]$_x$} for the same $(6,1,1)$ magnon mode assuming $M_{S}^{\textrm{YIG}}$ has been somehow reduced  to $1750$~G. We obtain  \red{${\mathcal C}_{\textrm{YIG}} \gtrsim 1$} for the smallest thin-film\cite{haidar2015thickness} $\alpha$ and \red{${\mathcal C}_{\textrm{YIG}}\approx 90\approx 6{\mathcal C}_{\textrm{V[TCNE]}_x}$} for  the $\alpha \approx 5\times 10^{-5}$ observed {\cite{Tabuchi2014,doi:10.1063/1.5115266}} in low-temperature bulk YIG
(see Supplemental Material). 
Therefore, we  {suggest} that despite the large $M_S$ of YIG (which enhances ${\mathcal C}_\lambda$ through an enhanced rf fringe field), the large observed thin-film damping values at low-temperature currently  make YIG a less favorable material than {V[TCNE]$_x$} for realizing the strong coupling regime. However, if  thin-film YIG damping values could be brought towards the bulk value $\alpha_{\textrm{YIG,bulk}}$~ {\cite{Tabuchi2014,doi:10.1063/1.5115266}}, and the resonance frequency mismatch fixed ({\it e.g.,} by reducing $M_S$ by about 25\% --  {keeping in mind that this would also imply \red{an} increase of the damping term}), the cooperativity could exceed that of {V[TCNE]$_x$} in the desired geometry (1~$\mu$m diameter disk directly on diamond).

{\it Experimental challenges.---} The experimental realization of this device structure presents a series of challenges that can be grouped into four broad categories: i) the ability to create {V[TCNE]$_x$}  {disks} on a diamond surface without loss of magnon coherence, ii) the creation of near-surface NV centers with long spin lifetime,  iii) measurement at milli-Kelvin temperatures, and iv) developing approaches to gate the interaction, thus enabling controlled entanglement. We briefly consider those challenges here.

Recent work has demonstrated that {V[TCNE]$_x$} can be patterned down to micron length scales on arbitrary substrates while maintaining damping values consistent with the thin-film values quoted here\cite{franson2019low} {($\alpha = 4\times 10^{-5}$)}. 
Prior results have shown that various packaging strategies allow for measurement under ambient conditions~\cite{Froning2015} and down to cryogenic temperatures \cite{Fang2011}. Fluctuations of paramagnetic spins at the diamond surface  {(with pre-placed or pre-identified NV centers)} have been shown to contribute to NV spin dephasing, and so in a real device some compromise must  be made between shallow positioning for increased coupling to magnons and deeper positioning for longer spin coherence times. Combining our calculations with recent experimental measurements of NV spin lifetime~\cite{nvdepth3} suggest that a depth of roughly 30 nm will provide the optimal cooperativity. Finally, measuring at temperatures below 100~mK presents experimental challenges, however, they have been solved previously for both microwave and optical measurements~\cite{Tabuchi2014}. The damping of {V[TCNE]$_x$} magnons at these temperatures has not been studied, however we note that in studies of YIG under similar conditions the \textit{intrinsic} magnon damping decreases with decreasing temperature. In YIG, this decrease is swamped by a dramatic increase in \textit{extrinsic} damping due to fluctuations of paramagnetic defects in the GGG substrate and in the YIG film itself~\cite{PhysRevLett.3.30,PhysRevLett.3.32,doi:10.1063/1.1984756,PhysRev.123.1937,PhysRev.133.A728,doi:10.1063/1.1984589}. We anticipate that the ability to deposit {V[TCNE]$_x$} directly onto the diamond surface will substantially mitigate these effects and may allow us to exploit the intrinsic decrease in damping. Finally it will be important to shift the bias magnetic field on microsecond timescales to bring the NV center spin(s) in and out of resonance with the magnon modes. This can be achieved with Oersted fields generated by passing currents through lithographically defined wires on the diamond surface, a geometry previously realized for [111]-oriented NV centers\cite{Fuchs2011}.

{\it Concluding remarks.---} The  spin-magnon coupling predicted here is suitable for entanglement and transduction among  NV spins and single magnon occupancy of a magnon mode.
Through an analytical calculation we find that for high angular index magnon modes of a 1~$\mu$m {V[TCNE]$_x$} disk,  the spin-magnon coupling strength \red{$\textsl{g} \approx 2\pi\times 10$}~kHz. We also predict a cooperativity of \red{${\mathcal C}_\lambda\approx 15$} within a wide spatial area within the diamond substrate, sufficiently far below the surface to avoid uncontrolled surface spin  noise, thus making our proposal experimentally realizable and not sensitively dependent on the  NV center position. We also calculate ${\mathcal C}_\lambda$ for YIG instead of {V[TCNE]$_x$}, and we find smaller  cooperativities in the case of thin-film YIG, due to the larger demagnetization field and the \red{larger} low-temperature, patterned thin-film damping. The geometry and materials we propose here provide a practical approach for strongly coupling   NV centers that are separated  by $\approx 1$~$\mu$m, providing possible avenues to assist quantum networking. 

\ack

The material is based on work supported by the U.S. Department of Energy, Office of Basic Energy Sciences, under Award Number DE-SC0019250. We would like to acknowledge D. D. Awschalom, M. Chilcote, A. R. da Cruz, K. Hu, S. R. McMillan, and C. \c{S}ahin for useful and valuable discussions. 


\begin{thebibliography}{10}
\bibitem{fringe}
The fringe fields of higher modes possess nodes along the vertical ($z$) and
  radial ($r$) directions, which in turn diminish the amplitude of the fringe
  fields.
\bibitem{temper}
Here we use the {V[TCNE]$_x$} damping $\alpha$ measured at room temperature.
\end{thebibliography}
\end{document}